\documentstyle[11pt,newpasp]{article}

\input psfig
 
\def\ha{H$\alpha$}

\begin{document}

\title{Distribution of Dark Matter in Bulge, Disk and Halo
inferred from High-accuracy Rotation Curves\footnote{ Proc. "Galaxy Dynamics:
from early times to the present", IAP Ap Meeting, 1999 July, 
ed. F.Combes, G.Mamon and V.Charmandaris, ASP Conf. Series}}
\author{Yoshiaki SOFUE}
\affil{Institute of Astronomy, University of Tokyo, Mitaka, Tokyo 181-8588}

\begin{abstract}

High-accuracy rotation curves of spiral galaxies show a steep rise in 
the central few hundred
pc region, indicating high concentration of mass toward the center.
Using the rotation curves, we idrectly calculate radial distributions 
of the surface-mass density (SMD), and obtain radial profiles of the 
mass-to-luminosity ratio (M/L). 
The M/L ratio is found to vary significantly 
not only in the massive halo but also within the bulge and disk regions.
It increases with the radius within the disk, and rapidly toward halo.
In some galaxies, the M/L increases inward toward the center 
within the bulge, indicating a massive dark core. On these bases, 
we discuss the radial 
distribution of dark matter in spiral galaxies. 
\end{abstract}


\section{Introduction}

The dark matter inferred from analyses of flat rotation curves 
dominates the mass of galaxies 
(Rubin et al 1980, 1982; Bosma 1981; Mathewson et al 1996; Persic et al 1996). 
However, the distribution of dark-matter in the inner disk and bulge
are not thoroughly investigated yet  because of the difficulty in 
observing rotation curves of the innermost part of galaxies.
In order to investigate the inner kinematics and rotation properties
of spiral galaxies, we have shown that the CO molecular line  
would be most useful because of its high concentration in the center 
as well as for its negligible extinction through nuclear gas disks
(Sofue 1996, 1997, Sofue et al 1997, 1998, 1999).
Recent high-dynamic range CCD spectroscopy in optical lines
has also made it possible to obtain high accuracy rotation curves for 
the inner regions (Rubin et al 1997; Sofue et al 1998). 
In this article, we review the observations of 
high-accuracy rotation curves of spiral galaxies, 
and discuss their general characteristics. 
Using the rotation curves, we derive the distribution of surface 
mass density (SMD), and discuss the radial 
variation of mass-to-luminosity ratio and the dark mass fraction. 

\section{Universal Properties of Rotation Curves}
 
Persic et al (1995,1996) have extensively studied the universal properties
of rotation curves with particular concern to the outer disk and
massive halo, who have used H$\alpha$ position-velocity data 
from Mathewson et al (1996).
However, inner rotation curves currently obtained by a method to
trace peak-intensity velocities on optical spectra may have missed 
higher velocity components in the central region, being affected by the
bright bulge light.
The inner kinematics is also not well traced by HI observations
because of the lack in HI  in the central regions (Bosma 1981a, b). 

We have proposed to use the CO-line emission in order to 
overcome this difficulty for its negligible extinction as well as its
high concentration toward the center.
Recent high dynamic-range CCD spectroscopy in the \ha\
and [NII] line emissions also provides us with accurate kinematics
of the central regions (Rubin et al 1998; Sofue et al 1998; 
Bertola et al 1998).
In deriving rotation curves, we applied the envelop-tracing
method from PV diagrams.

Fig. 1 shows well-sampled rotation 
curves obtained from these CO and CCD observations (Sofue et al 1999). 
Since the dynamical structure of a galaxy varies with the radius
rapidly toward the center, a logarithm plot
would help to overview the innermost kinematics.
In fact, logarithmic plots in Fig. 2 demonstrate the 
convenience to discuss the central kinematics. 
In such a plot, we may argue that high-mass galaxies show 
almost constant rotation velocities from the center to outer edge.
 
\begin{figure}
\psfig{figure=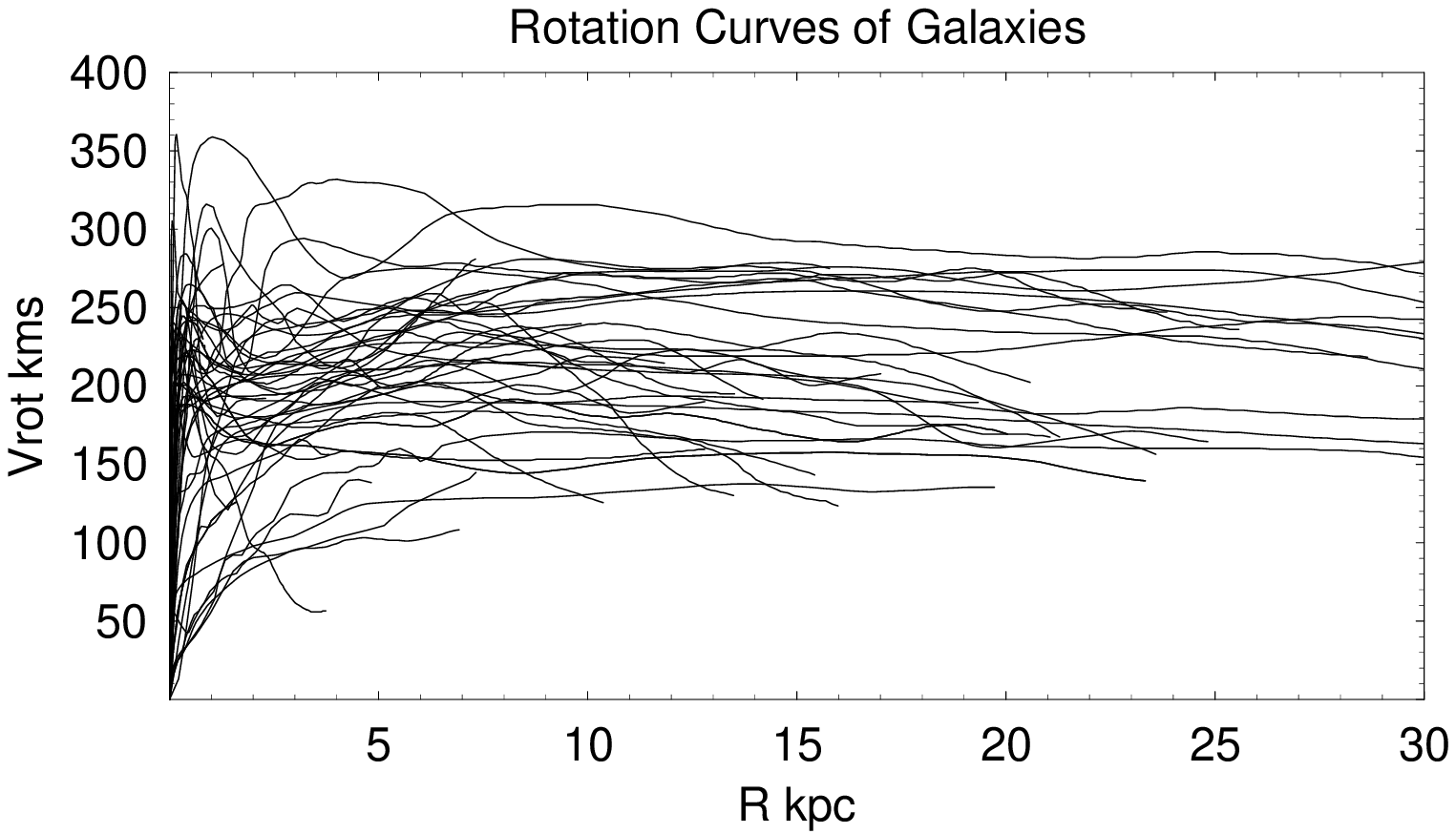,height=5cm} 
Fig. 1. High-accuracy rotation curves  of Sb, Sc, SBb 
and SBc galaxies obtained by using CO, \ha\ and HI-line data. 
\end{figure}
 
\begin{figure}
\psfig{figure=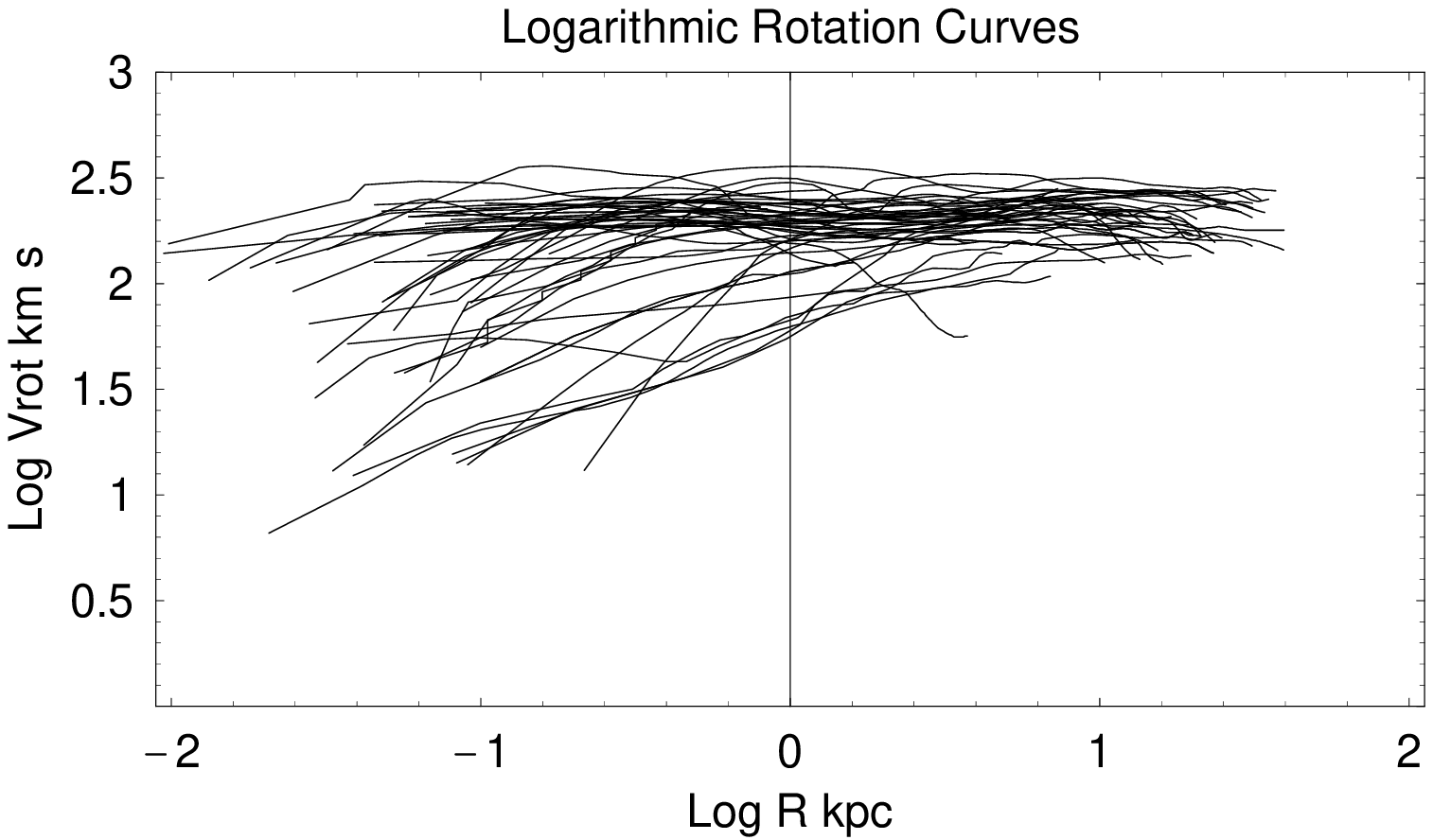,height=5cm} 
Fig. 2. Logarithmic rotation curves for the same galaxies as in Fig. 1. 
\end{figure}
 
We may summarize the universal properties of  rotation curves
in Fig. 1 and 2 as follows, which are similar to those for the Milky Way.
 
\vskip 2mm
{\parindent=0pt 
(1) Steep central rise: Massive galaxies show a very steep rise of 
rotation velocity in the central region. 
Rotation velocity is often found to be finite even in the nucleus.
Small-mass galaxies tend to show milder, rigid-body increase of rotation
velocity in the center.

(2) Sharp central peak and/or shoulder corresponding to the bulge. 

(3) Broad maximum in the disk; and

(4) Flat rotation toward the edge of the galaxy due to the massive halo. 
}
\vskip 2mm

The steep nuclear rise of rotation is a universal property for 
massive Sb and Sc galaxies, regardless the morphological types.
On the other hand, less massive galaxies show a rigid-body rise, except for
the Large Magellanic Cloud (see the next section).
The fact that almost all massive galaxies have a
steep rise indicates that it is not due to
non-circular motion by chance.
If there is a bar, the interstellar gas is shocked and dense gas is
bound to the shocked lane along the bar, while the gas in
high-velocity streaming motion is not observed to be in the molecular form.
Therefore, observed velocities in the CO lines should manifest the velocities
in the shocked lane, which is approximately equal to the pattern speed.
Hence, the molecular line observations may underestimate the rotation velocity.

\section{Surface-Mass Density}

The mass distribution in galaxies has been obtained by assuming constant
mass-to-luminosity ratios in individual components of the bulge and disk,
using the luminosity profiles (e.g. Kent 1987).
However, the M/L ratio may not be constant even in a single component,
such as due to color gradient as well as the variation of the 
dark-matter fraction.  
Forbes (1992) has obtained the mass distribution as a function of the
radius, where he has derived mean density within a certain radius 
(integral surface density).
We have developed a method to derive a differential surface mass
density as a function of the radius in order to compare the mass distribution
with observed profiles of the surface luminosity (Takamiya and Sofue 1999):

Once an accurate rotation curve of a galaxy is given, we are able to
calculate the surface-mass density directly from the rotation velocity, $V(r)$,
as the following (Takamiya and Sofue 1999).
We assume that the 'true' mass distribution in a real disk galaxy will 
be between two extreme cases; spherical and axisymmetric flat-disk 
distributions. 
For these two extreme cases, 
the SMD, $\sigma(R)$, is directly calculated as follows (Takamiya and Sofue
1999): 

{\it Spherical case}: The mass $M(r)$ inside the radius $r$ is given by
\begin{equation}
M(r)=\frac{r {V(r)}^{2}}{G},
\end{equation}
where $V(r)$ is the rotation velocity at $r$. Then the SMD
${\sigma}_{S}(R)$ at $R$ is calculated by,
\begin{eqnarray}
{\sigma}_{S}(R) & = & 2 \int\limits_0^{\infty} \rho (r) dz.
\end{eqnarray}
Here, $r=\sqrt{R^2+z^2}$ with $z$ the height from the galactic plane, and
the volume mass density $\rho(r)$ is given by
\begin{equation}
\rho(r) =\frac{1}{4 \pi r^2} \frac{dM(r)}{dr}.
\end{equation}
 
{\it Flat-disk case}: 
The SMD for a thin flat-disk, ${\sigma}_{D}(R)$, is derived by 
solving the Poisson's equation:
\begin{equation}
{\sigma}_{D}(R) =\frac{1}{{\pi}^2 G} \left[ \frac{1}{R} \int\limits_0^R 
{\left(\frac{dV^2}{dr} \right)}_x K \left(\frac{x}{R}\right)dx + 
\int\limits_R^{\infty} {\left(\frac{dV^2}{dr} \right)}_x K \left
(\frac{R}{x}\right) \frac{dx}{x} \right],
\end{equation}
where $K$ is the complete elliptic integral and becomes very large
when $x\simeq R$(Binney \& Tremaine 1987). 

The accuracy of this method can be examined by calculating 
SMD for the spherical and flat-disk assumptions, and the `true' SMD
for a known model potential.
We have used the Miyamoto-Nagai (MN) potential (Miyamoto \& Nagai 1975).  
Fig. 3 shows the true SMD and calculated SMDs for a spherical
and a flat-disk assumptions for an RC corresponding to the MN potential. 
The spherical case well reproduces the true SMD for the inner region. 
This is because a spherical component is dominant within the bulge. 
On the other hand, the flat-disk mimics the true one in the disk region, 
which is also reasonable. Near to the outer edge, the flat-disk case 
appears to be a better tracer of the true SMD. We stress that the 
results from for spherical and flat-disk assumptions do not differ by 
more than a factor of 1.5 from the true SMD, except for the outermost
region where the edge effect for the spherical case becomes  significant. 

\begin{figure}
\psfig{figure=fig3.ps,height=7.5cm} 
Fig. 3.  The real (solid line) SMD, and spherical (dashed),
and flat-disk (dot-dash) SMD profiles calculated for the rotation
curve of the Miyamoto-Nagai potential (lower panel)(Takamiya and Sofue 1999).
\end{figure}

\begin{figure}
\psfig{figure=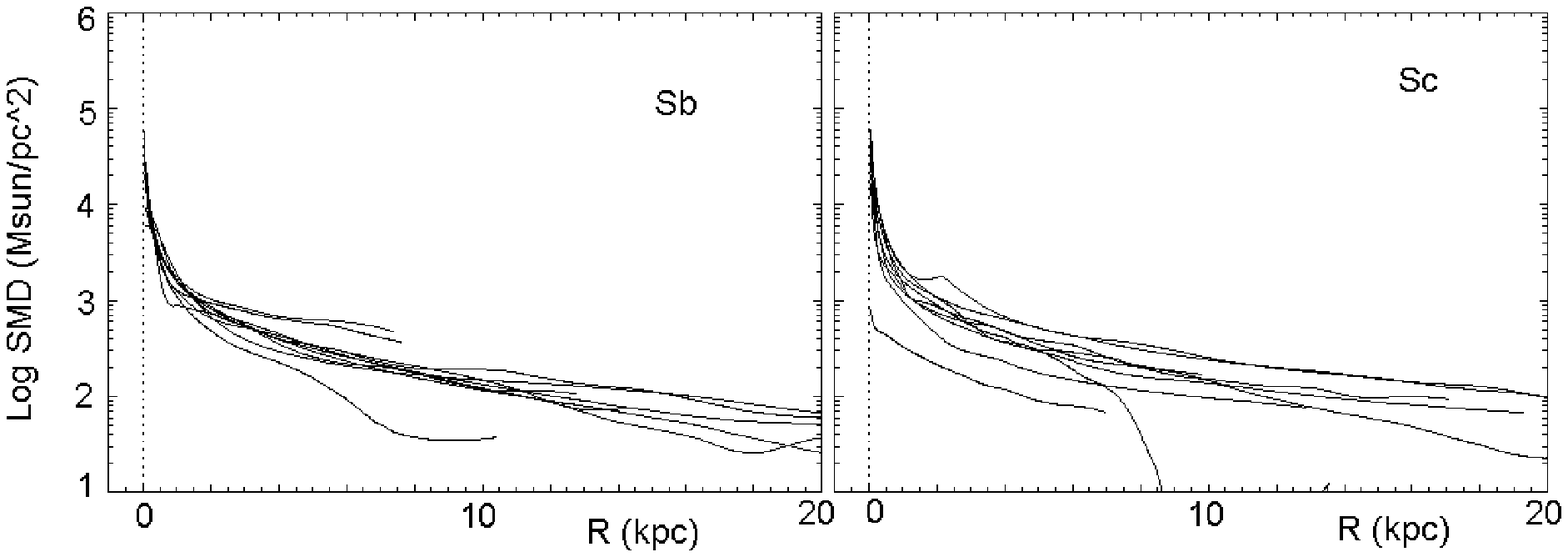,height=4.5cm} 
Fig. 4.  Surface-mass density profile for Sb and Sc galaxies (Takamiya and
Sofue 1999). 
\vskip 1mm

\psfig{figure=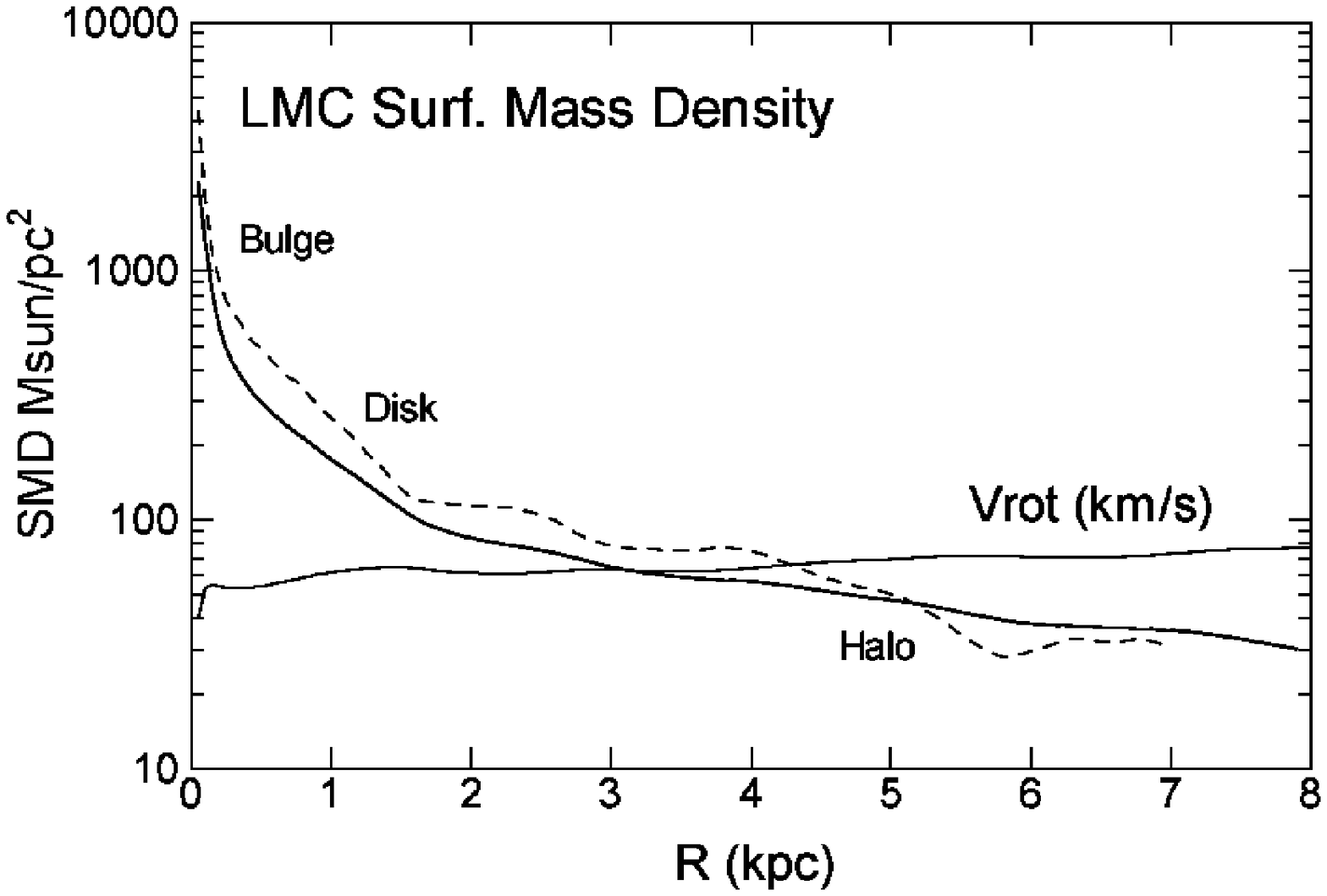,height=4cm} 
Fig. 5.  Radial profile of the SMD around the kinematical center of
the Large Magellanic Cloud, indicating a compact bulge component.
This 'dark bulge' is displaced by about 1.2  kpc from the optical bar center.
\end{figure}

We have calculated SMD profiles for Sb and Sc galaxies, for which
accurate rotation curves are available, and show the results in Fig. 4. 
Sb and Sc galaxies show similar surface-mass distributions to each other. 
In the disk region at radii 3 to 10 kpc the SMD decreases 
exponentially outward. 
In the bulge region of some galaxies, the SMD shows a power-law decrease, 
much steeper than an exponential decrease. 
This indicates that the surface-mass concentration in the central region 
is higher than the luminosity concentration. 
The central activities appear to be not directly correlated with the 
mass distribution. 
Note that the present sample includes galaxies showing Seyfert, LINER, 
jets and/or black holes.

Recently, high-resolution kinematics of the Large Magellanic Cloud 
in the HI line has been obtained by Kim et al (1998). 
Using their position-velocity diagrams, we have derived a rotation curve, 
and calculated an SMD distribution around the kinematical center.
Fig. 5 shows the thus obtained mass distribution in the LMC.
The SMD has a sharp peak near the kinematical center, indicating 
a dense massive core.
It is surrounded by an exponential disk and massive halo. 
The core component is not associated with any optical bulge-like 
component, and we called it a dark bulge.
The dark bulge is significantly displaced from the stellar bar.
Hence, such a dwarf galaxy as the LMC was found to show a similar mass-density
profile to the spiral galaxies as except for the absolute values.

\section{Mass-to-Luminosity Ratio and Dark-Matter Fraction}

\begin{figure} 
\psfig{figure=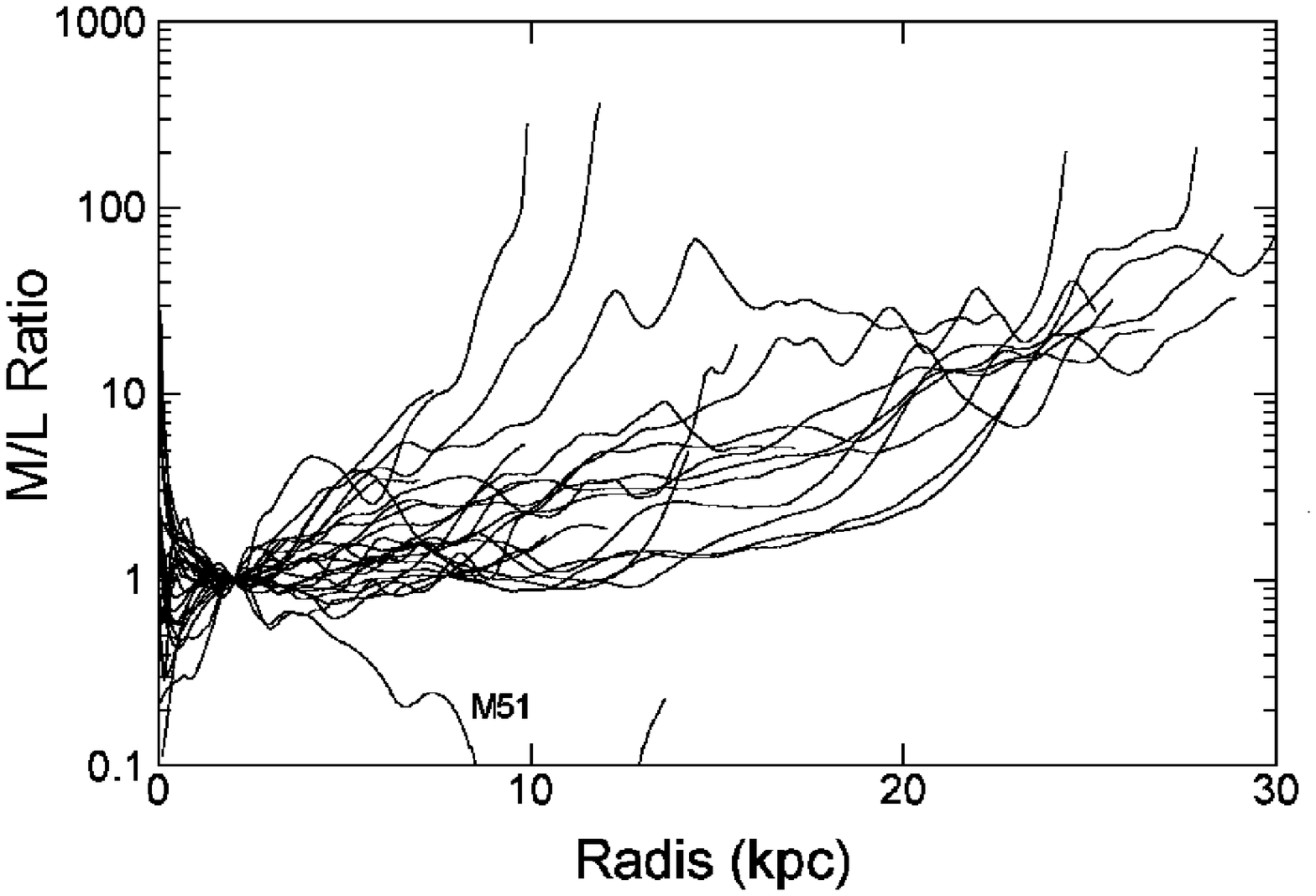,height=4cm} 
Fig. 6.  Mass-to-Luminosity ratio for Sb and Sc galaxies 
(Takamiya and Sofue 1999).
\vskip 1mm
\psfig{figure=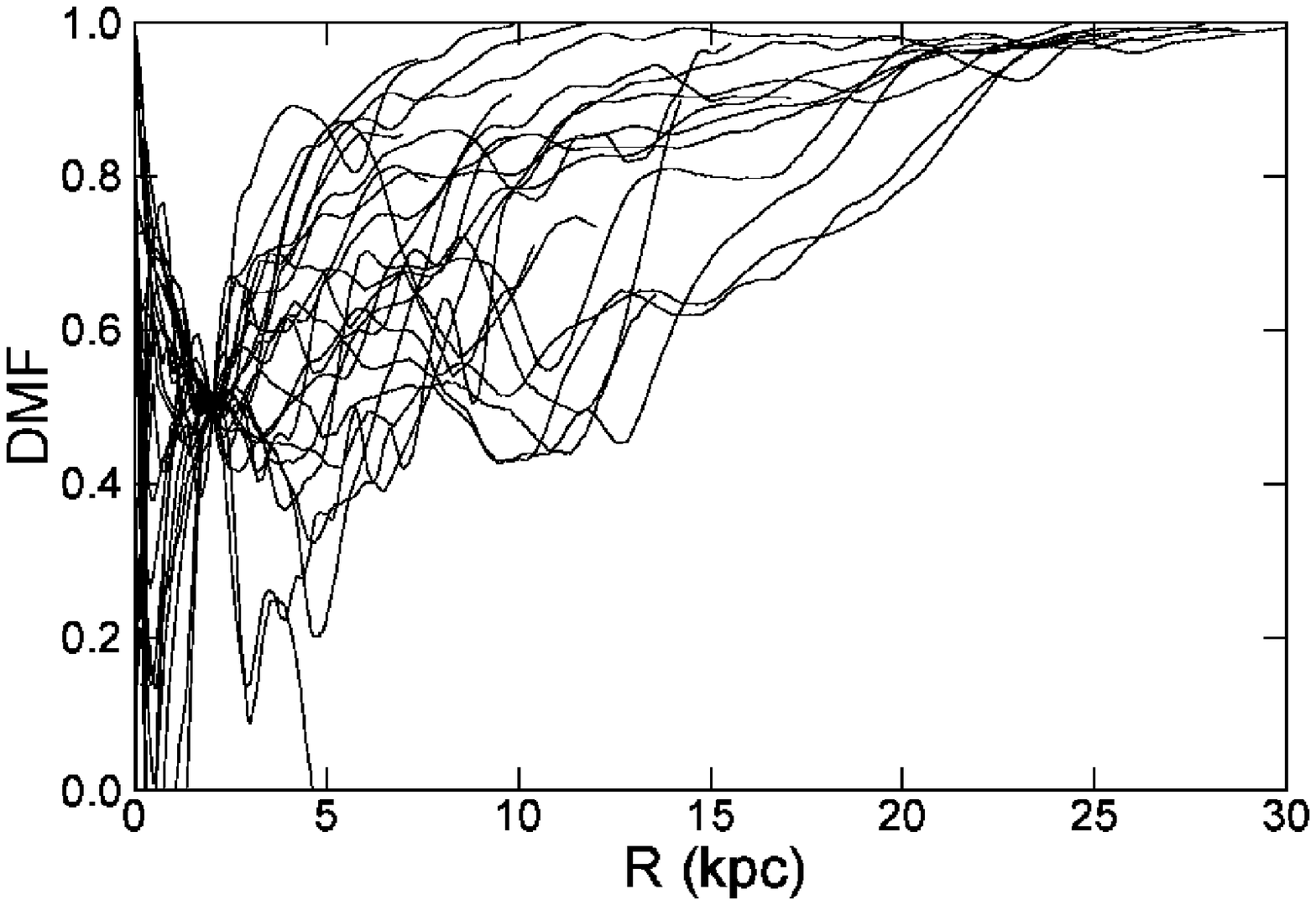,height=4cm} 
Fig. 7. Dark-Matter fraction, where DMF is fixed to 0.5 at $R=2$ kpc. 
This diagram shows a qualitative trend of the variation of dark matter
\end{figure}

The surface mass density can be, then, directly compared with
observed surface luminosity, from which we can derive the
mass-to-luminosity ratio (M/L) as normalized to unity at $R=2$ kpc.
Fig. 5 shows the thus obtained radial distributions of M/L 
for a disk assumption (Takamiya and Sofue 1999).
In order to see qualitatively the variation of dark-matter fraction,
we define a quantity, DMF, by
$$DMF= 1 - a/(M/L),$$
where $a$ is a constant and M/L is normalized to unity at $R=2$ kpc.
Here, we assume that the DMF is 0.5 at $R=2$kpc, or $a=0.5$.
Fig. 6 shows the thus obtained DMF profiles for the galaxies.

The figures indicate that the M/L and DMF is not constant at all,
but vary  significantly within a galaxy.  

(1) M/L and DMF vary drastically within the central bulge.
In some galaxies, it increases inward toward the center,
suggesting a dark massive core.
In some galaxies, it decreases toward the center, 
likely due to luminosity excess such as due to active nuclei.

(2) M/L and DMF gradually increases in the disk region, 
and the gradient increases with the radius.

(3) M/L and DMF increases drastically from the outer disk toward
the outer edge, indicating the massive dark halo.
In many galaxies, the dark halo can be nearly directly
seen from this figure, where the M/L exceed ten, and sometimes
hundred.  
fraction.

\section{Discussion}

 Although the variation of the M/L ratio  in the visible bands 
may partly be due to a gradient of stellar population, the large 
amplitude of the M/L variation may not entirely be attributed to the 
color gradient.  
The M/L and DMF profiles in Fig. 6 and 7 indicate that some galaxies 
(e.g., NGC 4527 and NGC 6946) show a very steep increase by more than an
order of magnitude toward the center within a few hundred pc radius.
Such variation of M/L is hard to be understood by color gradient in the
bulge. 
Such a steep central increase of M/L may imply that the bulge contains 
an excess of dark mass inside a few hundred pc, which we call a 
``massive dark core''. The scale radius is of the order of 100$\sim$200 pc,
and the mass is estimated to be $M \sim RV^{2}/G \sim 10^{9}M_{\odot}$. 
The massive dark 
cores may be an object linking the galactic bulge with a massive black 
hole in the nuclei (Miyoshi et al. 1995; Genzel et al. 1997; Ghez et al. 
1998) and/or massive core objects causing a central Keplerian RC 
(Bertola et al. 1998).

\end{document}